\documentclass[iop]{emulateapj}
\def\apj{{ApJ}}
\def\mnras{{ MNRAS}}

\usepackage{epsfig}
\usepackage{url}

\newcommand{\grb}{GRB~130427A}
\newcommand{\gr}{$\gamma$-ray}
\newcommand{\grs}{$\gamma$-rays}

\shorttitle{GeV emission of GRB~130427A}
\shortauthors{Tam et al.}

\begin{document}

\title{Discovery of an extra hard spectral component in the high-energy afterglow emission of GRB~130427A}

\author{Pak-Hin Thomas Tam\altaffilmark{1}, Qing-Wen Tang\altaffilmark{2,3}, Shu-Jin Hou\altaffilmark{4,5}, Ruo-Yu Liu\altaffilmark{2,3,6,7}, and Xiang-Yu Wang\altaffilmark{2,3}
}

\affil{$^1$ Institute of Astronomy and Department of Physics, National Tsing Hua University, Hsinchu 30013, Taiwan \\
$^2$ School of Astronomy and Space Science, Nanjing University, Nanjing, 210093, China \\
$^3$ Key laboratory of Modern Astronomy and Astrophysics (Nanjing University), Ministry of Education, Nanjing 210093, China \\
$^4$ Department of Astronomy and Institute of Theoretical
Physics and Astrophysics, Xiamen University, Xiamen, Fujian
361005, China \\
$^5$ Purple Mountain Observatory, Chinese Academy of Sciences,
Nanjing 210008, China \\
$^6$ Max-Planck-Institut f\"ur Kernphysik, D-69117 Heidelberg, Germany \\
$^7$ Fellow of the International Max Planck Research School for Astronomy and Cosmic Physics at the University of Heidelberg (IMPRS-HD)
}
\email{phtam@phys.nthu.edu.tw, xywang@nju.edu.cn}

\begin{abstract}
The extended high-energy gamma-ray ($>$ 100 MeV) emission occurred
after the prompt gamma-ray bursts (GRBs) is usually characterized by a
single power-law spectrum, which has been explained as the
afterglow synchrotron radiation. The afterglow inverse Compton
emission has long been predicted to be  able to produce a high-energy
component as well, but previous observations have not revealed
such a signature clearly, probably due to the small number of
$>$10~GeV photons even for the brightest GRBs known so far. In
this Letter, we report on the Fermi Large Area Telescope (LAT)
observations of the $>$100~MeV emission from the very bright and
nearby \grb. We characterize the time-resolved spectra of the GeV
emission from the GRB onset to the afterglow phase. By performing time-resolved spectral fits of \grb, we found a strong evidence of an extra hard spectral component that exists
in the extended high-energy emission of this GRB. We argue that
this hard component may arise from the afterglow inverse Compton
emission.

\end{abstract}


\section{Introduction}

One of the major scientific objectives of the Fermi Gamma-ray Space
Telescope  before launch is to characterize the spectrum of
gamma-ray bursts (GRBs) over 7 decades of photon energies, i.e.,
from 8~keV to over 100~GeV, both during the prompt emission phase
and the afterglow phase~\citep[see, e.g.,][]{lat_grb_prospects}.
Determining where a GRB spectrum ends at high energies has
important implications on GRB physics. The Fermi Large Area
Telescope (LAT) observations of GRBs over the past few years have
proved very fruitful, and results are summarized in the first LAT
GRB catalog and references therein~\citep{lat_grb_cat}. Major
results include the establishment of the  temporally extended GeV
emission from many LAT-detected GRBs.

The spectrum of the extended
emission above 100~MeV is usually characterized by a single power law. However,
due to the steeply decreasing photon flux from GRBs above 10 GeV,
the LAT has by now only detected a handful of $>$10~GeV photons
from GRBs; the most energetic of all observed photons from GRBs is
the $\sim$33~GeV photon emitted 82~s after the Fermi's Gamma-ray Burst Monitor (GBM) trigger of
GRB~090902B~\citep{lat_090902b}. An extra hard spectral component during the prompt phase has been detected in cases like GRB~090902B and GRB~090926A~\citep{lat_090902b,lat_090926a}. It is, however, yet unclear
whether the single power-law spectrum extends above 10 GeV or
there is any extra spectral components or even cutoffs in the GeV emission well after the prompt phase. This lack
of information is mostly due to the limited photons statistics
above a few GeV, where a single power-law spectrum seems to
adequately describe the $>$100~MeV spectra observed so far.

The widely discussed scenario for this temporally extended GeV
emission is the afterglow synchrotron model, where electrons are
accelerated by external forward shocks and produce GeV photons via
synchrotron radiation  (e.g. Kumar \& Barniol Duran 2009, 2010;
Ghisellini et al. 2009; Wang et al. 2010). This model works well
for high-energy emission above 100 MeV. However, since the
synchrotron radiation has a maximum photon energy
(typically $\la 50$ MeV in the rest-frame of the shock), it is
hard to explain the $>$10 GeV photons detected during the afterglow
phase (Piran \& Nakar 2010; Barniol Duran \& Kumar 2011; Sagi \&
Nakar 2012; Wang et al. 2013). Recently, \citet{wang_10gev}
suggested that these $>10$ GeV photons detected from some LAT GRBs
could originate from the synchrotron self-Compton (SSC) emission of the
early afterglow if the circum-burst density is sufficiently large,
or from external inverse Compton processes in the presence of
central X-ray emission~\citep{wang06}. The afterglow inverse Compton scenario has
been proposed for a while and has been extensively studied~\citep{meszaros94,zhang2001,Sari_Esin01,fan_08}.  The external inverse
Compton scenario (e.g. scattering off X-ray flare photons)  has
been supported by the simultaneous detections of X-ray flares (by
Swift) and GeV emission (by Fermi LAT) in GRB\,100728A
\citep{lat_100728a,He2012,zhang_lat_sample}.

Recently, based on similar theoretical arguments, \citet{fan_130427a}
suggested that inverse Compton scattering plays a crucial role in
giving rise to the $>$10~GeV photons from \grb~that arrive $>$100~s
after the GRB onset. In this Letter, we make use of the publicly
available LAT data to derive the basic characteristics of the
GeV emission from \grb. We derive the time-resolved spectra
during the afterglow phase and to study the presence of any
inverse Compton spectral component.


\section{Properties of \grb}
\grb~triggered several space instruments:
Swift/BAT~\citep{Maselli2013},  Fermi Gamma-Ray Telescope
\citep{Zhu2013a,Kienlin2013}, Konus-Wind \citep{Golenetskii2013},
SPI-ACS/INTEGRAL~\citep{Pozanenko2013}, and AGILE
\citep{Verrecchia2013}. Three RAPTOR full-sky persistent optical
monitors also recorded the event~\citep{Wren2013}. We adopt the
GBM trigger time, i.e., 07:47:06.42 UT on April 27, 2013, as
$T_\mathrm{0}$ throughout this work~\citep{Kienlin2013}.

The burst duration, $T_{90}$,
is about 138s as measured by the Gamma-Ray Burst Monitor (GBM) onboard Fermi in the energy range 50--300~keV.
Most notably, the energy fluence as measured in the prompt
emission phase is around 2$\times$10$^{-3}$~erg~cm$^{-2}$, putting
\grb~as the GRB with the highest fluence in the mission life of
both Fermi/GBM and Konus-Wind
\citep{Kienlin2013,Golenetskii2013}, as well as the
burst having the highest fluence measured by the LAT during the prompt phase.

\emph{Swift}'s X-ray Telescope (XRT) began data-taking of  the
burst at $T_\mathrm{0}+203$s and found the X-ray afterglow at the
position R.A.~$=11\mathrm{^h}32\mathrm{^m}32\fs63$,
Dec.~$=+27\arcdeg41\arcmin51\farcs7$~(J2000), with an error circle
of radius~$3\farcs5$~\citep[90\% confidence
level;][]{Kennea_gcn14485}. This position is used in the analyses
presented in this Letter. The flux faded steeply initially as a
power law with an index of $\alpha_1\sim2.8$ and, after a break at
$\sim T_\mathrm{0}+480$s, as
$\alpha_2\sim1.2$~\citep{evan_gcn14502}. The redshift of the burst
was found to be $z=0.34$~\citep{Flores2013,Levan2013,Xu2013}. At
this distance, its isotropic energy $E_{\rm \gamma,iso}$ is
$7.8\times 10^{53}$ erg~\citep{Kann_gcn14580}, making it the most
energetic GRB yet detected at $z\leq 0.5$.

The angle of the GRB position is about 47$^\circ$ from the  LAT
boresight when GBM was triggered and the GRB remains within the
LAT field-of-view (FoV) until around $T_\mathrm{0}+700$s~\citep{Zhu2013a}.
The GeV emission can be detected up to about one day after the
burst, although the GRB position had been occulted by the Earth
several times over such period~\citep{Zhu2013b}.

The LAT emission from \grb~lasts well beyond the prompt
emission period until  about one day after the GRB onset. This is
the longest GeV afterglow emission ever recorded for a GRB.
Furthermore, the GeV photons that \grb~emits contain some of the
most energetic \grs~from GRBs. A total of twelve $>$10~GeV photons
were recorded in the first 700~s after the burst onset, including
a 95.3~GeV photons arriving at $T_\mathrm{0}+243$s. At $z=0.34$,
this photon has an intrinsic photon energy (energy as measured at
the source frame) of 128~GeV, the highest ever known from a GRB.
Such a long duration and high photon energy may simply due to the proximity of \grb, as
compared to other bright LAT GRBs, since the observed
energy of a photon is the intrinsic photon energy divided by a
factor of $(1+z)$, and the late GeV afterglow of a far-away
\grb-like could be below the sensitivity level of the LAT. The
photon-photon attenuation of multi-GeV photons by the
extragalactic background light would only strengthen this
distance-only argument. On the other hand, the unique physical
conditions in the jet and/or the environment of \grb~might still
play a role in giving rise to the large number of $>$10~GeV
photons and the very high photon energy that they possess.

\section{LAT data analysis and results}
\subsection{Spectral analysis}
\label{spectra} GRB spectra are expected to change over time, so
we  performed time-resolved spectral analysis of the LAT data. The
Fermi Science Tools v9r27p1 package was used to reduce and analyze
the data between 100~MeV and 100~GeV. Using the
``P7TRANSIENT''-class data would increase the effective collection
area, and thus the photon statistics, by $\sim$50--100\% above
100~MeV, compared to the event class
``P7SOURCE\_V6''~\citep{lat_p7_instruments}.
Nevertheless, since our focus is in the extended GeV emission that
lasts for about one day, for simplicity we made use of the events classified as
``P7SOURCE'' for all analyses. The instrument
response functions ``P7SOURCE\_V6'' were used. To reduce the
contamination from Earth albedo $\gamma$-rays, we excluded events
with zenith angles greater than 100$^\circ$.

We then performed unbinned maximum-likelihood analyzes (\emph{gtlike}) of a 20$\degr$-ROI centered at the Swift/XRT position to characterize the spectra of the $>$100~MeV \grs~from the GRB onset to the afterglow phase. The Galactic diffuse emission (gal\_2yearp7v6\_v0.fits) and the isotropic diffuse component (iso\_p7v6source.txt), as well as sources in the second Fermi catalog were included in the background model. However, an isotropic component turns out to be enough to model the background photons in the time bins before $T_\mathrm{0}+1000$s, due to the dominance of the GRB emission over other sources in the ROI during these short-duration intervals.

\begin{figure*}
    \epsscale{.7}
   \plotone{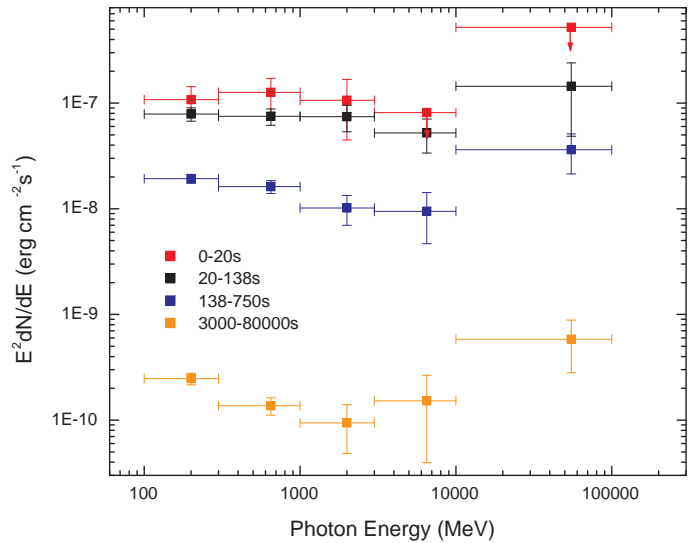}
      \caption{The 100~MeV to 100~GeV spectra from the prompt emission phase to the afterglow phase, spanning a total of 80~ks. Each horizontal bar represents the energy range used to derive the corresponding flux level (for that energy bin).}
         \label{five_band_spec}
   \end{figure*}

Firstly, we performed analysis for the whole energy range (i.e., 100~MeV to 100~GeV), assuming a single power law spectrum for the LAT photons from the GRB. Four time intervals from the prompt to afterglow phases were used. We found that the photon indices, $\Gamma\sim-2$ for all time bins (see Table~\ref{lat_spec}). However, it is expected that GeV spectra might change over time in the $\sim$1-day period. We therefore divided the LAT energy range into five energy bins, and performed likelihood analyzes for each bin, letting the photon index of the GRB to be free in each energy bin. We plot the flux given by the analysis of each bin in Fig.~\ref{five_band_spec}. It is apparent that the 100~MeV to 100~GeV spectra are indeed not well described by single power laws in some time intervals. 

We therefore employed the broken power-law model (BPL). The results are shown in Table~\ref{lat_spec}. It was found that the BPL fits the data better than a single PL at a significance level of $\sim$2.5--2.9 for the latter two time intervals (i.e., after 138~s), consistent with a visual inspection of Fig.~\ref{five_band_spec}. Fitting the data from 138s to 80ks with a BPL gives an improvement over a PL at a significance level of 3.5. We hence conclude that a soft ($\Gamma<-2$), low-energy spectral component below 1~GeV co-exists with a hard ($\Gamma\sim-1.4$) component at energies higher than a few GeV after $T_\mathrm{0}+138$~s . It is possible that the high-energy hard component already exists at the latter phase of the prompt emission, but our analysis found that a single power law fits the data as good as a broken power law in the second time interval, i.e., 20--138~s. An attempt to fit the data using the smooth broken power law, which contain an additional free parameter, did not provide constraining spectral fits.

\begin{deluxetable*}{cccccccc}
\tablecolumns{8}
\tabletypesize{\scriptsize}
\tablecaption{Spectral properties of the GeV emission for different time intervals. \label{lat_spec}}
\tablewidth{0pt}
\tablehead{
\colhead{$t-T_\mathrm{0}$ } & \colhead{ Power Law (PL)} & \multicolumn{3}{c}{Broken Power Law (BPL)} &\colhead{Improvement of BPL over PL\tablenotemark{a}} \\
 \cline{3-5} \\
\colhead{(sec)} & \colhead{$\Gamma$} & \colhead{$\Gamma_\mathrm{1}$ ($E<E_\mathrm{b}$)} & \colhead{$\Gamma_\mathrm{2}$ ($E>E_\mathrm{b}$)} & \colhead{$E_\mathrm{b}$ (GeV)} & \colhead{($\sigma$)}
}
\startdata
    0--20 & $-$2.0$\pm$0.2 & & \nodata & & \nodata \\
    20--138 & $-$1.9$\pm$0.1 & & \nodata & & \nodata \\
    138--750 & $-$2.1$\pm$0.1 & $-$2.2$\pm$0.1 & $-$1.4$\pm$0.2 & 4.3$\pm$2.0 & 2.5 \\
    3000--80,000 & $-$2.1$\pm$0.1 & $-$2.6$\pm$0.7 & $-$1.4$\pm$0.2 & 1.1$\pm$0.9 & 2.9 \\
    138--80,000 & $-$2.1$\pm$0.1 & $-$2.3$\pm$0.2 & $-$1.4$\pm$0.1 & 2.5$\pm$1.1 & 3.5
\enddata
\tablenotetext{a}{~calculated as $\sqrt{2\times[\log(\mathcal{L}_\mathrm{BPL})-\log(\mathcal{L}_\mathrm{PL})]}$}
\end{deluxetable*}


\subsection{Energy dependent light curves}
\label{energy_curve}

\begin{figure*}
    \epsscale{.7}
   \plotone{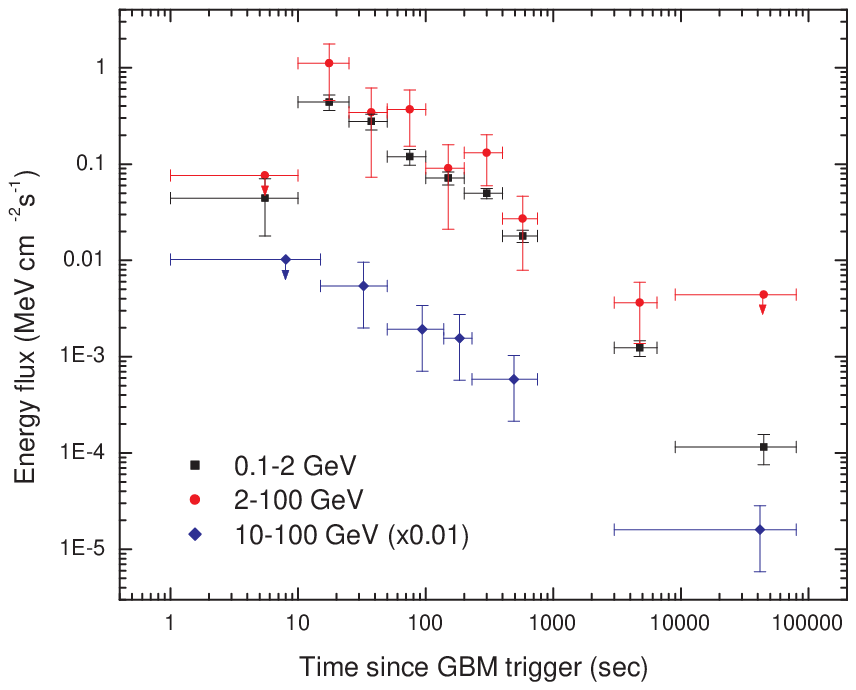}
      \caption{Time evolution of the energy flux from \grb~derived from the LAT data in three energy ranges: 100~MeV to 2~GeV (black squares), 2 to 100~GeV (red circles), and 10 to 100~GeV (blue diamonds). Upper limits are presented at the 90\% confidence level. The upper limits are derived assuming $\Gamma=-2.8$ (extrapolating from the $<$2~GeV spectrum in the same time interval) and $\Gamma=-1.4$ (see Sect.~\ref{energy_curve}), for the first and last time interval of the 2--100~GeV light curve, respectively. The $>$10~GeV light curve is scaled down by a factor of 100 for visualization purpose.}
         \label{energy_flux}
   \end{figure*}

To further investigate the time evolution of the two spectral components, we generated two light curves using unbinned likelihood analyses using photons below and above 2~GeV, respectively, as shown in Fig.~\ref{energy_flux}. The dividing line at 2~GeV was chosen to be roughly equal to the break energy, $E_\mathrm{b}$, found in the BPL model after $T_\mathrm{0}+138$~s (c.f. Table~\ref{lat_spec}). The time bins were chosen to ensure enough photon statistics in both bands (especially in the 2--100~GeV band). The low-energy (0.1--2~GeV) light curve can be described by a single power law decay with an index $\alpha_\mathrm{0.1-2\,GeV}=-1.1\pm0.1$, while the high-energy (2--100~GeV) light curve can be described by a single power law decay with an index $\alpha_\mathrm{2-100\,GeV}=-1.0\pm0.1$. 

During the last time range, i.e., 9~ks--80~ks after the burst, the likelihood analysis at energies 2--100~GeV formally returns a test-statistic value of 28 \citep[corresponding to a detection significance of about 5;][]{Mattox_96}, but the LAT only detected two photons above 2~GeV at that time, one of them having an energy 32~GeV, so we conservatively put an upper limit here. Assuming a photon index of $-$1.4 as derived from the broken-power fit in the 3~ks--80~ks time interval (see Sect.~\ref{spectra}), the upper limit of the energy flux $>$2~GeV during 9~ks--80~ks is the one shown in Fig.~\ref{energy_flux}. If we assume a very steep $\Gamma=-$2.8 (i.e., extrapolating the PL index found in a spectral analysis of the 0.1--2~GeV range during 9~ks--80~ks), the upper limit would become 3.4$\times$10$^{-4}$~MeV~cm$^{-2}$s$^{-1}$. However, we believe the spectral index is more likely to be hard based on the spectral studies in Sect.~\ref{spectra}. On the other hand, the detection of the single 32~GeV photon corresponds to an energy flux of $\sim$9$\times$10$^{-5}$~MeV~cm$^{-2}$s$^{-1}$.

\subsection{The $>$10 GeV photons}

Even compared to the four brightest LAT GRBs (GRB~080916C, GRB~090510, GRB~090902B, and GRB~090926A) known so far, \grb~is still peculiar in that it radiated over a hundred $>$1~GeV photons and a dozen photons at energies above 10~GeV. We dedicate this section to the $>$10 GeV photons emitted from \grb.

We present a list of $>$10~GeV photons associated with \grb~during the prompt and afterglow phases. First we selected all photons within an ROI of 3$^\circ$ centered on \grb~in the 80~ks after the burst. Following the formulation of \citet{kerr11}, a probability, $P_\mathrm{grb}$, that a photon is associated with \grb~(instead of coming from the background) was assigned to each photon, assuming the simple power law model as shown in Table~\ref{lat_spec}. The arrival times and energies of those photons of energy $>$10~GeV with $P_\mathrm{grb}>$99.5\% are shown in Table~\ref{10gev_photon}. We note that the energy resolution of 10--100~GeV photons (i.e. 68\% containment of the reconstructed incoming photon energy) is on the order of 10\%~\citep{lat_p7_instruments}.

We also produced a light curve using $>$10~GeV photons only. Due to the small number of $>$10~GeV photons, we follows the formulation of \citet{feldman98} to derive confidence levels based on the number of detected photons above 10~GeV in each time interval. Given the smaller photon statistics at these energies, longer time intervals are used, as compared to those in the light curves at lower energies as given in Sect.~\ref{energy_curve}. Over the past four and a half years, the number of $>$10~GeV photons that was detected from an ROI of radius one degree (which is roughly the point-spread-function of a 10~GeV photon) centered on the GRB position is only 11. The expected number of background photons is therefore very close to zero in short time ranges. The exposure was then calculated using the tool \emph{gtexposure} which is not sensitive to the assumed photon index. A large uncertainty is related to the conversion of the number of photons to the energy it corresponds, in which one has to assume a photon index at $>$10~GeV. Based on the results from spectral analysis, we assume $\Gamma=-2.0$ for those time bins before 230s, and $\Gamma=-1.4$ for those after 230s. We note that the energy flux thus calculated is increased by a factor of 2.3 when $\Gamma$ is changed from $-$2.0 to $-$1.4.

The 10--100~GeV light curve, as plotted in Fig.~\ref{energy_flux}, can be described by a single power law decay with an index $\alpha_\mathrm{10-100\,GeV}=-0.8\pm0.2$.  

\begin{table}
\centering
\caption{Properties of all $>$10~GeV photons associated with \grb. The energy resolution of 10--100~GeV photons is on the order of 10\%. \label{10gev_photon}}
\begin{tabular}{cccc}
    \hline\hline
    arrival time (since $T_\mathrm{0}$, in sec) & energy (GeV) \\
    \hline
    18.4 & 72.6 \\
    22.9 & 10.3 \\
    47.3 & 27.5 \\
    64.2 & 11.2 \\
    80.2 & 12.3 \\
    84.5 & 25.8 \\
    140.8 & 21.2 \\
    213.7 & 11.4 \\
    217.2 & 14.9 \\
    242.8 & 95.3 \\
    256.0 & 47.3 \\
    610.3 & 41.4 \\
    3409.6 & 38.5 \\
    6062.3 & 18.6 \\
    34365.9 & 32.0 \\
    \hline
\end{tabular}
\end{table}

\section{Discussion}

In the afterglow synchrotron emission scenario (Kumar \& Barniol
Duran 2009, 2010; Ghisellini et al. 2010; Wang et al. 2010), the
photon spectrum is characterized by a single power-law with
$\Gamma_{\rm LAT}=-(p+2)/2$  and the flux decays as
$t^{-(3p-2)/4}$, where $p$ is the electron distribution index. The
$0.1-2$ GeV emission in GRB130427A is well consistent with this
picture, taking $p=2.2$.

The hard spectral component ($\Gamma_{2}\ge -2.0$) above $E_{\rm
b}=4.3$ GeV  in GRB130427A should have another origin. A natural
scenario for this hard spectral component is the afterglow
SSC emission, in which a spectrum
$\Gamma=-(p+1)/2$ is expected above the peak frequency of the SSC
spectrum (i.e. $h\nu_m^{IC}<h\nu_{\rm obs}<h\nu_c^{IC}$, Sari \&
Esin 2001), where the two break frequencies are given by (Wang et
al. 2013)
\begin{equation}
h\nu_m^{IC}=0.1 {\rm GeV} f_p^4\epsilon_{e,-1}^4
\epsilon_{B,-5}^{1/2}E_{54}^{3/4}t_2^{-9/4}n_{0}^{-1/4}\\
\end{equation}
and
\begin{equation}
h\nu_c^{IC}=10^3 {\rm TeV}
\left(\frac{1+Y_c}{10}\right)^{-4}\epsilon_{B,-5}^{-7/2}E_{54}^{-5/4}n_{0}^{-9/4}t_2^{-1/4},
\end{equation}
$Y_c$ is the Compton parameter for electrons of energy $\gamma_c$
(i.e., the cooling Lorentz factor in the electron distribution)
and $f_p\equiv 6(p-2)/(p-1)$. Here $\epsilon_e$ and $\epsilon_B$
are, respectively, the equipartition factor for shock energy in
electrons and the magnetic field, $E$ is the blast wave energy, and
$n$ is the number density of the circum-burst medium. For $p=2.2$,
the expected photon index is $\Gamma_2=-1.6$, which is consistent
with the observed photon index $-1.4\pm0.2$ above the break. The
broken power-law spectrum may indicate that the SSC component is
dominant only at energies above several GeV and the synchrotron
component is dominant below that. This is possible since the
synchrotron radiation has a maximum photon energy, which is
thought to be below 10 GeV at hundreds of seconds after the burst
(Piran \& Nakar 2010; Sagi \& Nakar 2012; Lemoine 2013; Wang et
al. 2013).

The flux of the SSC component decreases as $t^{-(9p-11)/8}$ when
$\nu_m^{IC}$ falls below the observed frequency (Sari \& Esin 2001),
\begin{equation}
t_p\simeq 10 {\rm s} f_p^{16/9}\epsilon_{e,-1}^{16/9}
\epsilon_{B,-5}^{2/9}E_{54}^{1/3}n_{0}^{-1/9}(\frac{h\nu_{\rm
obs}}{\rm 10 GeV})^{-4/9}.
\end{equation}
So the flux above 10 GeV is expected to decay after the shock deceleration time $t_{\rm dec}=50\Gamma_{0,2.5}^{-8/3}E_{54}^{1/3}n_0^{-1/3}$s (where $\Gamma_0$
is the initial bulk Lorentz factor of the forward shock),  and with a slope of $t^{-1.1}$ (for $p=2.2$), which is consistent with the observed decay slope.

In this work, we have discovered an extra hard spectral component above a few GeV from \grb~that exists from $\sim$100~s up to one day after the GRB onset. This means that the afterglow spectrum of a GRB may extend to the very high energy \gr~range, i.e., $>$100~GeV. In fact, the redshift of \grb, $z\approx0.34$, puts it at a distance whose very high energy \grs~could have been detected~\citep{Xue09}.
GRB observations using current or future generations of the Imaging Atmosphere Cherenkov Telescopes (IACTs), such as H.E.S.S. II, MAGIC II, VERITAS, and CTA, not only minutes but hours after the prompt emission phase  is thus crucial to study such a hard spectral component in more details, thanks to their much larger effective collection area than the LAT.

 \acknowledgments
We thank the referee for comments and a rapid reply. This project is supported by the National Science Council of the
Republic of China (Taiwan)  through grant
NSC101-2112-M-007-022-MY3,  the 973 program under grant
2009CB824800, the NSFC under grants 11273016,  10973008, and
11033002, and the Excellent Youth Foundation of Jiangsu Province
(BK2012011). PHT would like to thank the hospitality of The University of Hong Kong, where this manuscript was written.


\begin{thebibliography}{}

\bibitem[Abdo et al.(2009)]{lat_090902b} Abdo, A. A. et al. (Fermi/LAT collaboration)\ 2009, \apjl, 706, L138
\bibitem[Abdo et al.(2010)]{lat_090510} Abdo, A. A. et al. (Fermi/LAT collaboration)\ 2010, \apj, 716, 1178
\bibitem[Abdo et al.(2011)]{lat_100728a} Abdo, A. A. et al. (Fermi/LAT collaboration) 2011, \apjl, 734, L27
\bibitem[Ackermann et al.(2011)]{lat_090926a} Ackermann, M. et al. (Fermi/LAT collaboration)\ 2011, \apj, 729, 114
\bibitem[Ackermann et al.(2012)]{lat_p7_instruments} Ackermann, M. et al. (Fermi/LAT collaboration) 2012, \apjs, 203, 4
\bibitem[Ackermann et al.(2013a)]{lat_grb_cat} Ackermann, M. et al. (Fermi/LAT collaboration) 2013a, preprint[arXiv:1303.2908]
\bibitem[Ackermann et al.(2013b)]{lat_110731a} Ackermann, M. et al. (Fermi/LAT collaboration) 2013b, \apj, 763, 71
\bibitem[Band et al.(2009)]{lat_grb_prospects} Band, D.~L., Axelsson, M.,  Baldini, L., et al.\ 2009, \apj, 701, 1673
\bibitem[Evan et al.(2013)]{evan_gcn14502} Evans, P.~A., Page, K.~L., Maselli, A., Mangano, V., Capalbi, M., Burrows, D.~N.\ 2013, GCN Circ. 14502
\bibitem[Fan et al.(2008)]{fan_08} Fan, Y.-Z., Piran, T.,  Narayan, R., \& Wei, D.-M.\ 2008, \mnras, 384, 1483
\bibitem[Fan et al.(2013)]{fan_130427a} Fan, Y.-Z. et al. 2013, submitted, preprint[arXiv:1305.1211]
\bibitem[Feldman \& Cousins(1998)]{feldman98} Feldman, G. J. \& Cousins, R. D. 1998, \prd, 57, 3873
\bibitem[Flores et al. (2013)]{Flores2013} Flores, H. et al., 2013, GCN Circ. 14491
\bibitem[Ghisellini et al.(2010)]{Ghisellini10} Ghisellini, G., Ghirlanda, G., Nava, L., \& Celotti, A.\ 2010, \mnras, 403, 926
\bibitem[Golenetskii et al. (2013)]{Golenetskii2013} Golenetskii, S., et asl. 2013, GCN Circ. 14487
\bibitem[He et al. (2012)]{He2012} He, H.-N., Wu, Zhang, B.-B., Wang, X. Y., Li, Z. \& M\'{e}sz\'{a}ros, P. 2012, ApJ, 753, 178
\bibitem[Kann \& Schulze(2013)]{Kann_gcn14580} Kann, D.~A. \& Schulze, S.\ 2013, GCN Circ. 14580
\bibitem[Kennea et al.(2013)]{Kennea_gcn14485} Kennea, J.~A.\ 2013, GCN Circ. 14485
\bibitem[Kerr(2011)]{kerr11} Kerr, M. 2011, \apj, 732, 38
\bibitem[von Kienlin et al. (2013)]{Kienlin2013} von Kienlin, A., 2013, GCN Circ. 14473
\bibitem[Kumar \& Barniol Duran(2009)]{BarniolDuran09} Kumar, P. \& Barniol Duran, R.\ 2009, \mnras, 400, L75
\bibitem[Kumar \& Barniol Duran(2010)]{BarniolDuran10} Kumar, P. \& Barniol Duran, R.\ 2010, \mnras, 409, 226
\bibitem[Lemoine(2013)]{Lemoine13} Lemoine, M.\ 2013, \mnras, 428, 845
\bibitem[Levan et al. (2013)]{Levan2013} Levan, A. J., Cenko, S. B., Perley, D. A., \& Tanvir, N. R. 2013, GCN Circ. 14455
\bibitem[Maselli et al. (2013)]{Maselli2013} Maselli, A., Beardmore, A. P., Lien, A. Y., Mangano, V., Mountford, C. J., Page, K. L., Palmer, D. M., \&  Siegel, M. H. 2013, GCN Circ. 14448
\bibitem[Mattox et al.(1996)]{Mattox_96} Mattox, J.~R., et al.\ 1996, \apj, 461, 396
\bibitem[M\'esz\'aros \& Rees(1994)]{meszaros94} M\'esz\'aros, P. \& Rees, M.\ 1994, MNRAS, 269, L41
\bibitem[De Pasquale et al.(2010)]{090510_afterglow} De Pasquale, M., et al. 2010, \apjl, 709, L146
\bibitem[Piran \& Nakar(2010)]{Piran_Nakar10} Piran, T., \& Nakar, E.\ 2010, \apjl, 718, L63
\bibitem[Pozanenko et al.(2013)]{Pozanenko2013} Pozanenko, A., Minaev, P., \& Volnova, A. 2013, GCN Circ. 14484
\bibitem[Sagi \& Nakar(2012)]{Sagi12} Sagi, E., \& Nakar, E.\ 2012, \apj, 749, 80
\bibitem[Sari \& Esin(2001)]{Sari_Esin01} Sari, R. \& Esin, A.~A.\ 2001, \apj, 548, 787
\bibitem[Tam et al.(2012)]{lat_110625a} Tam, P.~H.~T., Kong, A.~K.~H., \& Fan, Y.-Z. 2012, \apj, 754, 117
\bibitem[Verrecchia et al. (2013)]{Verrecchia2013} Verrecchia, F., et al. 2013, GCN Circ. 14515
\bibitem[Wang et al.(2006)]{wang06} Wang, X. Y., Li, Z., \& M\'esz\'aros,~P. 2006, \apjl, 641, L89
\bibitem[Wang et al.(2010)]{wang_KN_effect} Wang, X.-Y., He, H.-N., Li, Z., Wu, X.-F., \& Dai, Z.-G.\ 2010, \apj, 712, 1232
\bibitem[Wang et al.(2013)]{wang_10gev} Wang, X.-Y., Liu, R.-Y., \& Lemoine, M.\ 2013, arXiv:1305.1494
\bibitem[Wren et al. (2013)]{Wren2013} Wren, J., Vestrand, W. T., Wozniak, P., \& Davis, H. 2013, GCN Circ. 14476
\bibitem[Xu et al. (2013)]{Xu2013} Xu, D., et al. 2013, GCN Circ. 14478
\bibitem[Xue et al. (2009)]{Xue09} Xue, R. R., Tam, P. H., Wagner, S. J., Behera, B., Fan, Y. Z., Wei, \& D. M. 2009, ApJ, 703, 60
\bibitem[Zhang \&  M\'esz\'aros(2001)]{zhang2001} Zhang, B., \&  M\'esz\'aros, P., 2001, ApJ, 559, 110
\bibitem[Zhang et al.(2011)]{zhang_lat_sample} Zhang, B.-B., et al.\ 2011, \apj, 730, 141
\bibitem[Zhu et al. (2013a)]{Zhu2013a} Zhu, S., Racusin, J., Kocevski, D., McEnery, J., Longo, F., Chiang, J., \& Vianello, G. 2013a, GCN Circ. 14471
\bibitem[Zhu et al. (2013b)]{Zhu2013b} Zhu, S., Racusin, J., Kocevski, D., McEnery, J., Longo, F., Chiang, J., \& Vianello, G. 2013b, GCN Circ. 14508
\bibitem[Zou et al.(2009)]{Zou09} Zou, Y.-C., Fan, Y.-Z., \& Piran, T.\ 2009, \mnras, 396, 1163

\end{thebibliography}
\end{document}